\documentclass[twocolumn]{aastex631}
\usepackage{booktabs}

\shorttitle{Thermal OH Emission in M31}
\shortauthors{Busch}

\begin{document}

\title{First Extragalactic Detection of Thermal Hydroxyl (OH) 18cm Emission in M31 Reveals Abundant CO-faint Molecular Gas}

\author[0000-0003-4961-6511]{Michael P. Busch}
\affiliation{Department of Astronomy \& Astrophysics, 
University of California, San Diego,
9500 Gilman Drive,
San Diego, CA 92093, USA}
\correspondingauthor{Michael P. Busch}
\email{mpbusch@ucsd.edu}




\begin{abstract}
The most abundant interstellar molecule, molecular Hydrogen (H$_{2}$), is practically invisible in cold molecular clouds. Astronomers typically use carbon monoxide (CO) to trace the bulk distribution and mass of H$_{2}$ in our galaxy and many others. CO observations alone fail to trace a massive component of molecular gas known as ``CO-dark'' gas. We present an ultra sensitive pilot search for the 18cm hydroxyl (OH) lines in the Andromeda Galaxy (M31) with the 100m Robert C. Byrd Green Bank Telescope. We successfully detected the 1667 and 1665 MHz OH in faint emission. The 1665/1667 MHz line ratio is consistent with the characteristic 5:9 ratio associated with local thermodynamic equilibrium (LTE). To our knowledge, this is the first detection of non-maser 18cm OH emission in another galaxy. We compare our OH and HI observations with archival CO (1-0) observations. Our OH detection position overlaps with the previously discovered Arp Outer Arm in CO. Our best estimates show that the amount of H$_{2}$ traced by OH is 140\% higher than the amount traced by CO in this sightline. We show that the amount of dark molecular gas implied by dust data supports this conclusion. We conclude that the 18cm OH lines hold promise as a valuable tool for mapping of the ``CO-dark'' and ``CO-faint'' molecular gas phase in nearby galaxies, especially with upcoming multi-beam, phased-array feed receivers on radio telescopes which will allow for drastically improved mapping speeds of faint signals.
\end{abstract}



\section{Introduction} \label{sec:intro}

The hydroxyl radical (OH) was the first molecule discovered in the radio regime in absorption towards Cas A 60 years ago \citep{Weinreb1963RadioMedium}. There was initial optimism at the time that OH would serve as a tracer for the bulk of H$_{2}$ in the galaxy. It was quickly discovered that this would be incredibly difficult due to the complex excitation of the four $\Lambda$-doubling ground-state lines (1612, 1665, 1667 and 1720 MHz), and the faintness of the emission signal \citep{Barrett1967RadioRadicals}. Surveys for OH in our own Galaxy revealed that absorption measurements were far more successful than emission due to the typically low excitation temperature (a few K above background continuum and CMB) of the  18cm OH lines. The first major survey for OH was carried out by \citet{Goss1968OHGalaxy} in absorption. Shortly after the discovery of OH 18 cm absorption, \citet{Weaver1965ObservationsMolecule} announced the detection of OH 18 cm emission. Surveys for OH emission from the OH 18cm lines tended to be relatively faint and was difficult to detect using the receiver technology of the time \citep{Penzias1964}, although thermal (or ``normal'') emission from OH was eventually announced \citep{Heiles1968NormalClouds}. The discovery of the masing phenomenon of OH molecules captured attention as a brighter target for observational OH work \citep{Wilson1968DiscoveryStars,Baan1987}. Meanwhile, millimetre emission from the CO molecule, first detected by \citet{Wilson1970CarbonNebula} took over as the choice tracer for H$_{2}$ in the Galaxy, owing to its brightness and abundance \citep[for a review see e.g.][]{Bolatto2013TheFactor, Heyer2015MolecularWay}.

Attempts to detect OH emission in external galaxies have historically struggled with the weak strength of the signal and limitations of receiver technology at the time. The first systematic search for OH in 63 nearby spiral galaxies was led by \citet{Schmelz1988AGalaxies} using the 305m Arecibo Observatory. There was an early search for OH in the Large Magellanic Cloud (LMC) using the Parkes 64m Telescope \citep{Radhakrishnan1967}. The detection of OH maser emission from OH/IR stars in the LMC \citep{Wood1986} prompted additional searches for OH masers in the Magellanic clouds \citep{Wood1992}. The now well known and incredibly bright OH megamaser phenomenon was first discovered in Arp 220 \citep{Baan1987}, and was recently revisited with much higher angular resolution data \citep{Baan2023TheStory}. A large absorption survey of OH in the Magellanic systems is expected to be carried out by GASKAP this decade \citep{Dickey2013}. A recent search for OH absorption using the FAST telescope by \citet{Zheng2020} failed to detect any extragalactic OH absorption sources, but placed stringent upper limits, and also reviewed the existing literature on extragalactic OH absorption sources. Of note, one of the earliest searches for extragalactic OH using the Effelsberg 100m telescope successfully observed thermal OH in absorption in M82 \citep{Nguyen-Q-Rieu1976OH82}. More recently, masing 1720 MHz OH emission was detected in a galaxy at $z$ = 0.247 in the MeerKAT Absorption Line Survey \citep{Combes2021PKSSpectrum}.

The recent renewed interest in searching for OH emission is partly due to the utility of OH to trace the ``CO-dark'' molecular gas. It was originally discovered by \citet{Wannier1993WarmObservations} that OH emission can arise at cloud edges, where no corresponding CO emission was detected. The observational discovery of ``dark gas'', a massive diffuse ISM gas phase seemingly not traced by HI or CO, but implied to exist by indirect total gas tracers like $\gamma$-rays and dust \citep{Grenier2005UnveilingNeighborhood, Wolfire2010TheGas, PlanckCollaboration2011PlanckGalaxy, Murray2018OpticallyISM} prompted others to consider the OH emission as a tracer for a diffuse molecular gas phase not traced by CO. The first modern and sensitive blind search for OH emission from the quiescent ISM by \citet{Allen2012Faint5circ} demonstrated that OH emission was ubiquitous, and commonly not accompanied by CO emission. A wealth of research has also demonstrated the existence of ``CO-dark'' molecular gas: the C[II]$\lambda$158um line has been used to trace the ``CO-dark'' H$_{2}$ \citep{Pineda2013AComponents, Madden2020, Chevance2020TheDoradus,  Bigiel2020, Schneider2023IonizedClouds}, simulations of molecular cloud formation and photodissociation region modeling on cloud and galactic scales \citep{Seifried2019SILCC-Zoom:Fields, Nickerson2019TowardsGalaxy, Inoguchi2020FactoriesFeedback, skalidis2022}, and molecular absorption of other tracers such as CH \citep{Jacob2019FingerprintingDeconvolution, Jacob2021TheLines, Jacob2022HyGAL:IRS1}, HF \citep{kavak2019}, and HCO$^{+}$ \citep{liszt2019,Rybarczyk2022TheNOEMA, liszt2023}.

\begin{figure*}
    \centering
    \includegraphics[width=\textwidth]{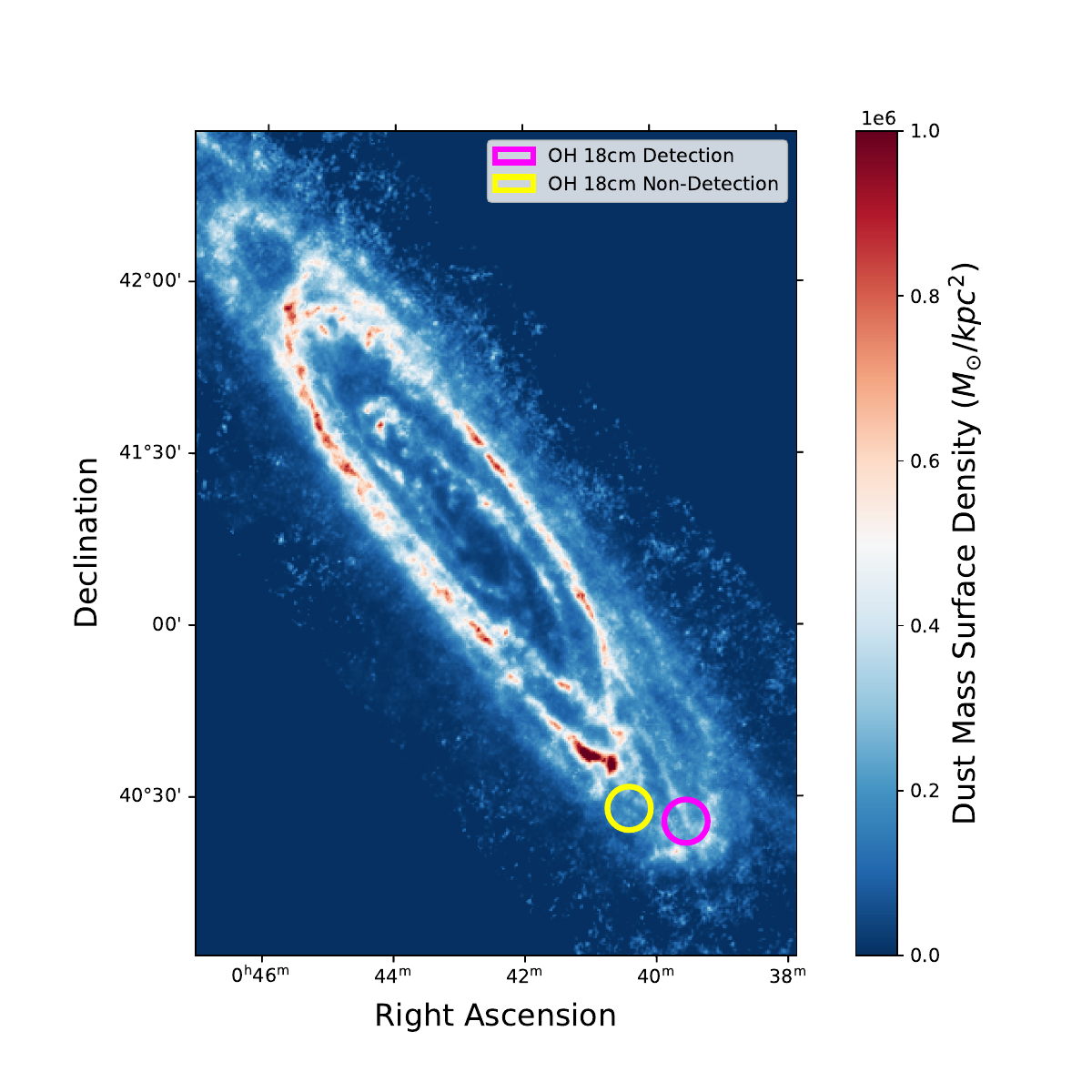}
    \caption{The derived mass surface density map of dust from \citet{Draine2014AndromedasDust}. The spectroscopic measurement of 18cm OH discussed in this paper is shown at (J2000) $0^{h}39^{m}32^{s}$$+40^{\circ}26'$ as a magenta open circle, approximately 62' from the nucleus, aligned with the major axis of the galaxy. The yellow open circle is the sightline with a non-detection at (J2000) $0^{h}40^{m}24^{s}$$+40^{\circ}28'$.}
    \label{fig:coordinate}
\end{figure*}

The SPLASH survey mapped OH in the galactic plane with the Parkes 64m telescope and greatly improved our understanding of the distribution of OH in the Galaxy \citep{Dawson2014SPLASH:Region}. OH was widely observed in all four ground state transitions. The OH profiles remain difficult to interpret because of the low excitation temperatures and non-uniformity and brightness of the continuum temperature towards the inner galaxy, leading to a mix of absorption and emission profiles along the line of sight. However, a large amount of observational knowledge concerning the anomalous excitation \citep{Petzler2020TheRegions, Petzler2021AMOEBA:Algorithm} and catalogs of OH masers have been generated \citep{Dawson2022SPLASH:Release}. 

The observational situation for OH is remarkably different towards the outer galaxy, where the ambient radio continuum temperatures are uniformly low (within 1K of T$_{CMB}$). Faint OH emission (T$_{mb}$ $<$ 0.05K) has been repeatedly detected in sensitive blind OH emission surveys, which are usually in the ``main-line'' LTE ratio (1665/1667, 5:9) \citep{Allen2012Faint5circ,Allen2015The+1deg,Busch2019}. The weaker ``satellite'' lines (1612, 1720 MHz, 1:1 ratio in LTE) are routinely too faint to detect. In reality, there are commonly departures from the LTE ratio in the satellite lines of interstellar OH. The 1720 MHz line can sometimes be slightly anomalously emitting \citep{Allen2015The+1deg}, even if the main-lines are in the LTE 5:9 ratio. This behaviour may be tracing shocks moving through the ISM leading to compression and increased collisional excitation \citep{Lockett1998OHIndicators}, or be indicative of the slightly anomalous ambient excitation temperatures of the satellite lines--as the SPLASH survey observed widespread diffuse weakly-masing 1720 MHz emission, even across entire clouds which are not near supernova remnants \citep{Dawson2014SPLASH:Region}. 

Altogether, OH has been demonstrated to trace the ``CO-dark'' H$_{2}$ in diffuse clouds \citep{Barriault2010MultiwavelengthEmission, Cotten2012HydroxylMBM40, Engelke2018OHW5, Engelke2019OHW5, Allen2015The+1deg, Busch2019}, in the envelopes of giant molecular clouds \citep{Wannier1993WarmObservations}, in absorption sightlines across the sky \citep{Li2015QuantifyingGas,Li2018WhereDMG,petzler2023}, and more recently in a thick ``CO-dark'' molecular gas ``disk'' of ultra-diffuse H$_{2}$ (n$_{H}$ $\sim$ 5 $\times$ 10$^{-3}$ cm$^{-3}$) in the outer galaxy \citep{Busch2021}. Potentially related, broad faint HCO$^{+}$ absorption coincident with HI absorption with no corresponding CO emission or absorption was also recently discovered in several sightlines \citep{Rybarczyk2022TheNOEMA}. HCO$^{+}$ and OH optical depths have been shown to be tightly correlated in diffuse gas regimes \citep{liszt1996}. A similar discovery of very broad Galactic OH emission and HCO+ absorption was presented back in 2010 by \cite{Liszt2010TheGas} towards one sightline.

We present here, to our knowledge, the first detection of thermal OH emission in another Galaxy, M31. We stress that by ``thermal'', we mean strictly that the 1665:1667 MHz emission is in the 5:9 LTE line ratio. The 1720 MHz line is undetected, which is consistent with the LTE ratio. As the 1720 MHz line would be nine times weaker than the 1667 MHz line with the LTE ratio, any such emission would be below our current sensitivity limits. The 1612 MHz line is not recovered because of radio frequency interference (RFI) at the Green Bank Observatory. There is no evidence that the emission is caused by unresolved masing sources in the beam, for which the line strengths would not be in the LTE ratio.

\section{Observations and Data} \label{sec:data}

\begin{table*}[t]
    \centering
    \caption{GBT L-band (1.15-1.73 GHz) Observing Parameters}\label{Observing Parameters}
    \begin{tabular}{cc}
    \toprule
    \midrule
      Project Code  & AGBT20A\_556  \\
      Observing Dates   & 2020-06-04 to 2020-07-21 \\
      Backend & VEGAS \citep{Prestage2015ThePlans} \\
      Polarizations & X,Y (linear) \\
      Spectral Windows & 4 \\
      Central Frequencies (MHz) & 1420.40, 1617.40, 1666.40, 1720.53 \\
      Pointing Source & 0114+4823 \\
      Bandwith (MHz) & 16.875 \\
      Spectral Resolution (KHz) & 1.03 \\
      Integration time per spectral dump (s) & 5 \\
      Beam Size (') & 7.6 \\
      \bottomrule
    \end{tabular}
\end{table*}

\subsection{M31 Blind Pilot Survey Construction}

We chose two positions in M31 to observe for OH 18 cm emission, at $\alpha$ = $0^{h}39^{m}32^{s}$, $\delta$ = $+40^{\circ}26'$ (J2000), and at $\alpha$ = $0^{h}40^{m}24^{s}$, $\delta$ = $+40^{\circ}28'$ (J2000). Both of these positions are approximately 62' from the nucleus, corresponding to a distance from the center of $\sim13$ kpc. This pilot survey was \textit{blind} in the sense that we did not point towards any previously known astronomical object, such as HII regions, supernova remnants, a potential absorption source, etc. The goal was to attempt to observe quiescent diffuse molecular gas on a large physical scale. We did however choose these sightlines for several other reasons. First, these sightlines contain relatively bright HI emission, which may increase the likelihood of detecting an OH emission signal at the same coordinates, since recent OH surveys in our Galaxy have demonstrated that HI and OH brightness temperatures are correlated, at least until the HI becomes optically thick \citep{Allen2012Faint5circ, Allen2013ERRATUM:97, Allen2015The+1deg, Busch2021}. Secondly, these two sightlines are spatially separated from the bulk of the CO emission in M31, between 8-11 kpc, as presented in \citep{Dame1993} and any OH detections could be indicative of ``CO-dark'' or ``CO-faint'' molecular gas at large galactic radii \citep{Wannier1993WarmObservations,Grenier2005UnveilingNeighborhood, Wolfire2010TheGas, Li2015QuantifyingGas, Li2018WhereDMG}, giving us the opportunity to learn more about molecular gas in regions of galaxies where there is an absence of, or very faint, CO emission. Third, we wanted to optimize the detectability of faint OH emission. The continuum temperature at radio wavelengths is typically quite low at large galactic radii, just above T$_{CMB}$. Generally, this is below the observed range of excitation temperatures of the OH lines ($\approx$ 4-10K) \citep{Li2018WhereDMG, Engelke2018OHW5, petzler2023}. Background continuum temperatures above this range could result in a mix of absorption and emission in the main OH lines, or no detections at all if the continuum temperatures become approximately equal to the OH excitation temperature.

After 21.26 hours of the exploratory search program, faint OH emission was significantly detected in the first planned sightline ($\alpha$ = $0^{h}39^{m}32^{s}$, $\delta$ = $+40^{\circ}26'$ (J2000), the magenta aperture in Fig.\ref{fig:coordinate}), which had the higher peak brightness temperature in HI (T$_{mb}$ = 30K). We then spent the rest of the observing time searching for OH in a nearby sightline ($\alpha$ = $0^{h}40^{m}24^{s}$, $\delta$ = $+40^{\circ}28'$ (J2000), the yellow circle in Fig. \ref{fig:coordinate}) with the hope of detecting OH in another position. The second sightline had a peak brightness temperature in HI of T$_{mb}$ = 13K. A total of 37.30 hours of integration time was spent on this sightline; there was no statistically significant detection of OH at 1665, 1667 or 1720 MHz. The lack of a secondary OH detection implies that the OH detection in the primary sightline is not an unknown instrumental effect of the telescope or receiver which might have introduced a spurious signal. The spectra of both sightlines are shown in Fig \ref{fig:data2}.

\subsection{Green Bank Telescope Observations}

Observations were made with the Robert C. Byrd Green Bank Telescope (GBT) located in West Virginia within the National Radio Quiet Zone, using the Gregorian receiver system operating in the frequency band 1.15-1.73 GHz ($L$-band). Observing parameters are summarized in Table \ref{Observing Parameters}. We observed for 21.26 hours at $\alpha$ = $0^{h}39^{m}32^{s}$, $\delta$ = $+40^{\circ}26'$ (J2000) and for 37.30 hours at $\alpha$ = $0^{h}40^{m}24^{s}$, $\delta$ = $+40^{\circ}28'$ (J2000). The observations took place from June 2020 to July 2020. The signal is fed to the control by an IF system. The signal from the IF was copied and directed at 4 sections of the GBT spectrometer, centered on 1420.40 MHz, 1617.40 MHz, 1666.40 MHz, and 1720.530 MHz. In order to maximize on-signal observing time we employed in-band frequency-switching by $\pm$2 MHz \citep{ONeil2002SingleWavelengths}, such that the expected science signal appears in both switching cycles, but in different channel numbers. The spectra from both switching cycles are therefore shifted and inverted before subtraction to average the science signal from both switching cycles. The IF band width was 16.875 MHz, chosen in order to minimize the possibility of harmful radio interference, and to make it possible to observe the OH main line 1665 MHz and 1667 MHz spectra on the same IF band. The frequencies observed also covered the OH satellite lines at 1612 MHz and 1720 MHz, although the 1612 MHz spectra suffer frequent transient baseline ripples from radio interference. We simultaneously observed the HI 21 cm line at 1420 MHz. Each pointing is made up of a sum of multiple 10-minute scans. The receiver is a single-beam, dual polarization instrument with an effective system temperature in the range of 16 - 20 K, depending on the weather, elevation of the pointing and background continuum emission. The FWHM of the GBT point-spread function at 18 cm is 7.6'. The antenna efficiency of the GBT in this frequency band is $\eta$ = 0.95, as determined by NRAO staff\footnote{\url{https://www.gb.nrao.edu/GBT/Performance/PlaningObservations.htm}}, hence the antenna and main-beam brightness temperatures are virtually identical ($T_{mb} = T_{A}/0.95$). The pointing of the telescope was calibrated at the start of each observing session. The GBT project number for the data presented in this paper is AGBT20A\_556.

\subsection{Archival CO Data}\label{Codata}

We compare our OH observations with archival CO (1-0) observations taken with the 1.2 m telescope at the Harvard Center for Astrophysics (CFA) \citep{Dame1993}. This is the only available CO survey of M31 that has overlapping data with our survey. This CO survey has observed all of M31 out to a radius of at least 15 kpc and has a similar beamsize at 8.7' to the GBT beam at OH (7.6'), and it is extremely sensitive, at an rms of 18mk per 1.3 km/s channel. The extraction position of the spectra is shown on the integrated CO map in Fig. \ref{fig:COmap}.

\subsection{Data Reduction}

The HI observations are checked for RFI in each 10 minute scan and then averaged together. A polynomial of order two is fit to 100 km/s on either side of the HI signal and subtracted. The data are smoothed by Gaussian convolution and decimated by 5 channels and the final velocity resolution is 1 km/s and the root-mean-squre noise per channel is 4 mK. For the OH observations, each 10 minute scan in each linear polarization is reviewed for the presence of interference or other instrumental problems. In general, there are no major sources of radio frequency interference (RFI) near the 1667, 1665 or 1720 MHz lines. Unfortunately there are large amounts of RFI near the 1612 MHz line, which render it unusable. 

Before smoothing and decimation, the noise is in the spectrum is approximately 5mK, the spectral resolution is 1.88 kHz corresponding to a velocity resolution of 0.34 km/s. The resultant noise per channel is 0.4 mK at the final chosen velocity smoothing of 3.88 km/s, which corresponds to a spectral resolution of 22.24 kHz. The noise levels agree very well with the radiometer equation for the integration times and channel bandwith, which is a good confirmation that we are not overfitting the background in the signal-free baseline regions.

After averaging all scans, the final spectra are smoothed by Gaussian convolution, with decimation, by 27 channels to a velocity resolution of 3.88 km/s to decrease rms noise per channel. An order 4 polynomial baseline was subtracted from the final baseline by fitting 100 km/s on either side of the expected signal. The boundaries for the baseline fitting followed the extent of the HI emission. This was done using the usual baseline fitting procedures in \textit{GBTIDL}, such at \textit{setregion}, \textit{nfit} and \textit{baseline} \citep{Garwood2006GBTIDL:Data}.

Initial inspection of the CO spectra at our detection position revealed a broad CO signal coincident with the OH and HI spectra. This was determined to be at least part of the previously discovered Arp Outer Arm S5 in CO in \citet{Dame1993}. We smoothed the CO signal with a convolution to a Gaussian kernel and decimated by 3 channels to a corresponding velocity resolution of 1.7 km/s using \textit{scipy} \citep{Virtanen2020SciPyPython}. This decreased the resulting rms in the spectra to 3mK per 1.7 km/s channel. The reprocessing confirmed the existence of broad, faint CO emission at the target position, with peak T$_{mb}$ of 17 $\pm$ 3 mK. The original CO data and the reprocessed data and Gaussian fit are shown in Fig. \ref{fig:CO}.

\begin{figure*}
    \centering
    \includegraphics[width=\textwidth]{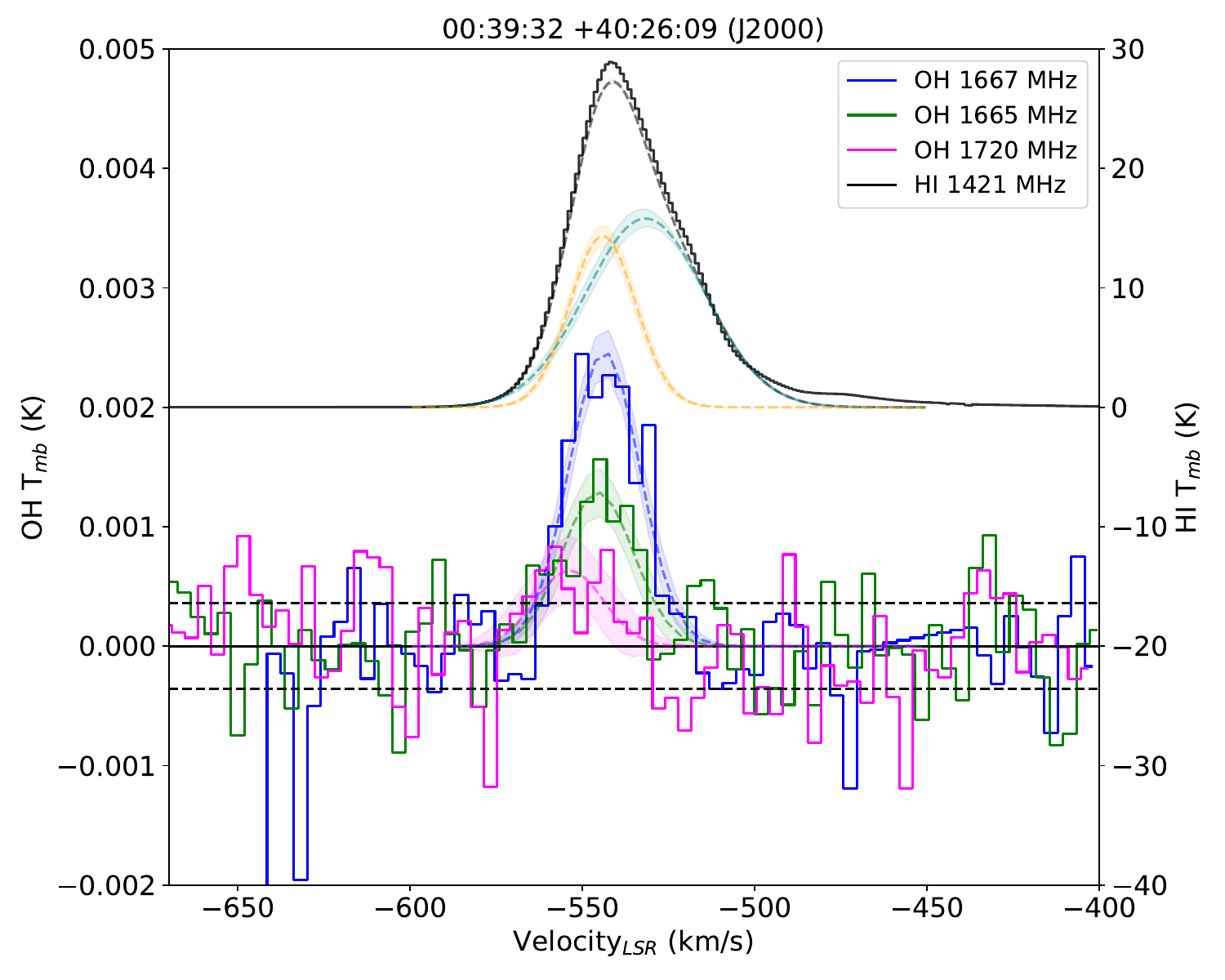}
    \caption{21cm HI and three 18cm OH lines overplotted. The scale on the left axis is for the OH data while the right is for HI. The OH spectra were smoothed to 3.88 km/s per channel to achieve necessary S/N, whereas the HI spectrum was smoothed to 1 km/s. The dashed line corresponds to $\pm$ 1-$\sigma$ statistical noise, $\Delta$ T$_{mb}$ $\sim$ 0.4mK. The steep drop off at the edge of the bandpass at -650 km/s in the 1667 MHz spectra is an artifact of baseline subtraction on the bandpass shape. Gaussian fits to each line are overplotted in dashed lines with a 1-$\sigma$ confidence band. In the case of the HI line, two Gaussian components were fit. The best fit parameters are reported in Table \ref{table:fits}.}
    \label{fig:data}
\end{figure*}

\section{Results} \label{sec:results}

Our most significant result is the detection of thermal 18cm OH emission in another galaxy for the first time, an optically thin radio wavelength tracer of H$_{2}$.

\subsection{Features of the HI and OH Line Profiles} \label{OH}

The observed HI and OH spectra from the GBT are displayed in Figure. \ref{fig:data}. The peak of the OH emission at 1667 MHz is 2.4 mK, whereas the peak of the 1665 MHz emission is is 1.5mK, close to the expected LTE ratio of the lines. The 1612 MHz line is unusable due to RFI at Green Bank. We do not detect any emission, or absorption, in the 1720 MHz line. This is expected for LTE conditions: if the 1720 MHz emission were in the LTE ratio (1:5:9:1 for 1612:1665:1667:1720 MHz), then the peak 1720 MHz emission would be buried in the noise, as T$_{mb,1720}$ = $T_{mb,1667}$/9 $\sim$ 0.3 mK. As the rms noise per channel is about 0.4 mK, it is consistent with the LTE interpretation for 1720 MHz emission to be absent at the sensitivity of the observations presented here. 

The statistical error in W(OH) is calculated as the square root of the number of channels (N$_{chan}$ $\sim$ 100) in the profile integral times the 1-$\sigma$ baseline rms. The rms is calculated by the \textit{stats} routine in \textit{GBTIDL} between -900 km/s and -600 km/s (where no signal is present). Typical rms values calculated this way are approximately 0.4 mK, with a slight dependence on the order of baseline polynomial subtracted.

We analyze the HI and OH profiles by fitting a Gaussian profile to each line to derive the following line parameters: peak brightness, line width, line center, and their corresponding statistical uncertainties. We performed the line integrals directly on the data which in practice is a sum of channels multiplied by the channel width in km/s:

\begin{equation}
    \mathrm{W(OH)} = \sum T_{mb} \times \Delta V
\end{equation}

The Gaussian profiles are fit to the data using \textit{lmfit} \citep{Newville2014LMFIT:Python} and the results are reported in Table \ref{table:fits}. In the case of the HI profile, two Gaussians are fit as it seems there is a narrow and broad component. The central velocity and velocity dispersion of the lines are overall similar to each other, indicating that the emission likely arises from gas that is associated with each other. We do note that the broad component of HI is offset from the peak of OH, whereas the the narrow component has the same central velocity of the OH. 

The integrated line emission values, W(OH), from -600 km/s to -500 km/s are 0.056 $\pm$ 0.008 K km/s and 0.039 $\pm$ 0.008 K km/s for the 1667 and 1665 MHz lines respectively. This corresponds to a S/N of 7 for W(OH$_{1667}$) and 5 for W(OH$_{1665}$). In LTE, the ratio between the brightness temperatures vary between 1.8 for optically thin conditions to 1 for infinite optical depths \citep{Tang2017}. The profile integrals of the OH 1667 and 1665 MHz lines appear to be consistent in the LTE ratio as we calculate R = $\int\ T_{67}\ dv / \int\ T_{65}\ dv$ = 1.4 $\pm$ 0.34.  

\begin{figure*}[t]
\begin{tabular}{cc}
    \includegraphics[width=0.45\textwidth]{M31OHBoth_Detection.pdf}
    &
    \includegraphics[width=0.45\textwidth]{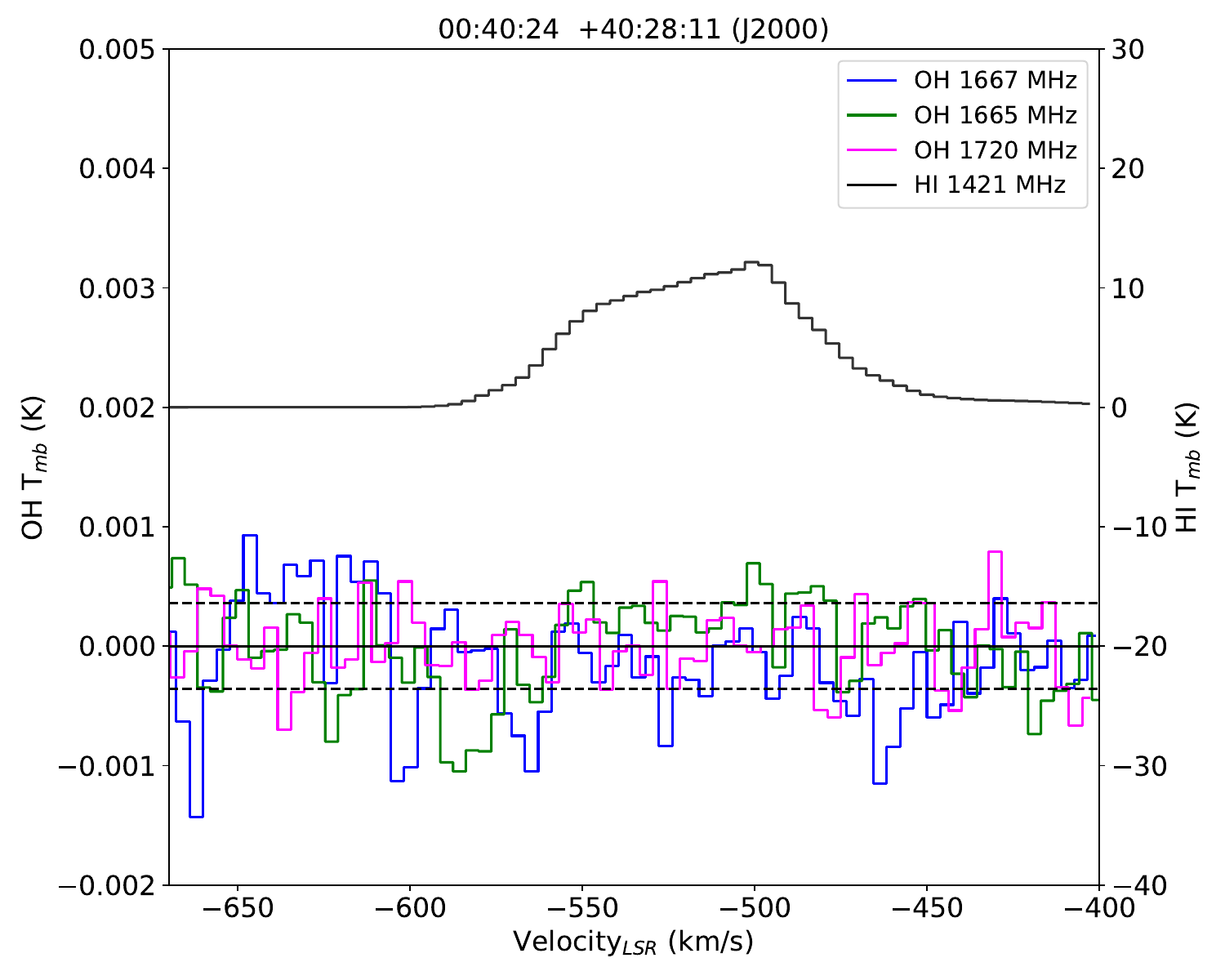}
    \end{tabular}
    \caption{Both sightlines discussed in this paper. OH was detected at the coordinates presented in the left panel. No significant OH signal was detected in the right panel (the 1667 MHz line is shown). Both OH spectra were reduced using the same procedures discussed in Section \ref{sec:data}. The observed GBT HI spectra for both positions are also overplotted. The dashed line corresponds to $\pm$ 1-$\sigma$ statistical noise, $\Delta$ T$_{mb}$ $\sim$ 0.4mK.}
    \label{fig:data2}
\end{figure*}

\subsection{Column Densities of HI and OH} \label{ColumnDensities}

The beam-averaged column density of HI can be calculated in the optically thin assumption by \citep[e.g.][]{Dickey1982NeutralInstrument}. The optically thin assumption leads to a lower limit for the HI column density. On the physical scale discussed in this paper, this is likely a safe assumption and we are not missing a large opaque fraction of HI. Under another assumption that the spin temperature is much larger than the continuum background at 21cm (T$_{s}$ $>>$ T$_{c}$), the column density can be calculated directly from the measured $T_{b}$:

\begin{equation}
    N(\mathrm{HI}) = C_{0}\
    \int \Delta T_b(v) dv ;
    \label{eqn:hi}
\end{equation}

where the constant $C_{0}$ is 1.82 $\times 10^{18}$ $\mathrm{cm^{-2}}$. The beam-averaged column density, $N$(OH), along the line of sight is calculated by \citep[e.g.][]{Liszt2010TheGas, Allen2012Faint5circ, Allen2015The+1deg, Busch2019}:

\begin{equation}
    N(\mathrm{OH}) = C_{0}\left[\frac{T_{ex}}{T_{ex} - T_c}\right]
    \int \Delta T_b(v) dv ;
    \label{eqn:oh}
\end{equation}

where $\Delta T_b(v)$ is the (main beam) brightness temperature of the OH emission profile from the cloud as observed in one of the 18-cm OH transitions, minus an estimate of the underlying radio continuum brightness at the same radial velocity, $T_{ex}$ is the excitation temperature for either the 1667 or 1665 line, $T_c$ is the brightness of the continuum emission at 1666 MHz incident on the back surface of the cloud, and the integration over velocity includes all the molecular emission thought to arise in that particular cloud. The prefactor, T$_{ex}$/T$_{ex}$-T$_{C}$ is usually called the 'background correction' factor, and a larger contrast between T$_{ex}$ and T$_{C}$ would lower the uncertainty in the column density correction. Unfortunately the excitation temperatures for the 18cm OH lines are usually seen to have excitation temperatures within a few kelvin of the background continuum temperature. The observed range is usually between 4-10K with the typical values being around 4-5K. \citep{liszt1996,Li2015QuantifyingGas,petzler2023}.

We require a reasonable measurement of the continuum temperature, T$_{c}$ in order to calculate the OH column density. The usual prescription is to find a continuum survey at a similar frequency, assume a spectral index and extrapolate to the continuum at 18cm \citep{Nguyen2018Dust-GasISM}. We use the 20.5cm (1462 MHz) Effelsburg+VLA continuum survey of M31 presented in \cite{Beck1998}. Inspecting their Figure 4 continuum map, we adopt an average intensity of 1 mJy/beam. Using the angular resolution of 45'', we transform this to temperature and then use the following equation to calculate T$_{C}$ at 18cm (1666 MHz) assuming a standard spectral index of -2.8 \citep{Nguyen2018Dust-GasISM}:

\begin{equation}
    T_{c,1666} = T_{CMB} + T_{C,1462}(1666/1462)^{-2.8}
\end{equation}

This results in a T$_{C}$ of 3.28 K. The rms noise in the continuum map near the map edges is stated to be $\approx$ 150 $\mu$Jy, and the corresponding error in T$_{C}$ is negligible. 

We have to assume an excitation temperature of the OH lines in the absence of emission/absorption pairs or further information. The distribution of T$_{ex}$ presented in \citet{Li2015QuantifyingGas}, which samples diffuse absorption sightlines in the Milky Way, is an appropriate starting point. We take T$_{ex}$ = 5K and 6K for 1667 and 1665 MHz lines respectively, as these lines have been shown to have different excitation temperatures by about 1K \citep{Engelke2018OHW5}. The limit in which T$_{ex}$ $>>$ T$_{C}$ is also a useful lower limit of $N$(OH). The constant $C_{0}$ is $2.257 \times 10^{14}$ cm$^{-2}$ for the 1667 MHz line. The value of $C_{0}$ for the 1665 MHz line is $4.0 \times 10^{14}$ cm$^{-2}$. 

We compute the HI column densities for both fit Gaussian components separately, we call broad and narrow corresponding to their fit velocity widths (FWHM of 40.6 and 22.6 km/s respectively). Using the above quantities, the computed HI column density is $N$(HI$_{broad}$) = 1.24 $\pm$ 0.01 $\times$ 10$^{21}$ cm$^{-2}$ and $N$(HI$_{narrow}$) = 6.26 $\pm$ 0.01 $\times$ 10$^{21}$ cm$^{-2}$, where the error is statistical. The column density of OH calculated from the 1667 MHz line is $N$(OH) = 3.7 $\pm$ 0.4 $\times$ 10$^{13}$. The column density calculated from the 1665 MHz line is $N$(OH) = 3.8 $\pm$ 0.7 $\times$ 10$^{13}$ cm$^{2}$, which is statistically consistent with the 1667 MHz derived column density. We calculate the abundance ratio of OH/HI as  6 $\times$ 10$^{-8}$, where we use the narrow component of the HI only.

\begin{table*}
    \centering
    \caption{Gaussian Fit Parameters and Profile Integrals Towards Source M31-540}
    \begin{tabular}{cccccc}
    \toprule
    Line Description & $v_{LSR}$ (km/s) & FWHM (km/s) & T$_{peak}$ (K) & $W$ (K km/s)& S/N \\ 
    \midrule
    HI 21cm (Narrow) & -544.09 $\pm$ 0.22 & 22.58 $\pm$ 0.77 & 14.32 $\pm$ 0.88 & 344 $\pm$ 32 & 10.75 \\
    HI 21cm (Broad) & -531.71 $\pm$ 0.71 & 40.61 $\pm$ 0.56 & 15.80 $\pm$ 0.72 & 683 $\pm$ 33 & 20.69 \\ 
    CO(J=1-0) & -548.44 $\pm$ 0.54 & 18.63 $\pm$ 1.28 & 0.017 $\pm$ 0.004 & 0.29 $\pm$ 0.02 & 14.5 \\ 
    OH 1667 MHz & -543.64 $\pm$ 0.94 & 24.15 $\pm$ 2.23 & 0.0024 $\pm$ 0.0004 & 0.057 $\pm$ 0.0077 & 7.3 \\ 
    OH 1665 MHz & -545.55 $\pm$ 1.84 & 24.54 $\pm$ 4.34 & 0.0014 $\pm$ 0.0004 & 0.040 $\pm$ 0.0079 & 5.7 \\ 
    OH 1720 MHz & -554.71 $\pm$ 4.80 & 22.77 $\pm$ 11.32 & 0.0009 $\pm$ 0.0004 & 0.00059 $\pm$ 0.0078 & 0.076 \\
    \bottomrule
    \label{table:fits}
    \end{tabular}
    \label{table:fits}
\end{table*}

\section{Discussion} \label{sec:discussion}

\begin{figure*}[t]
    \centering
    \includegraphics[width=\textwidth]{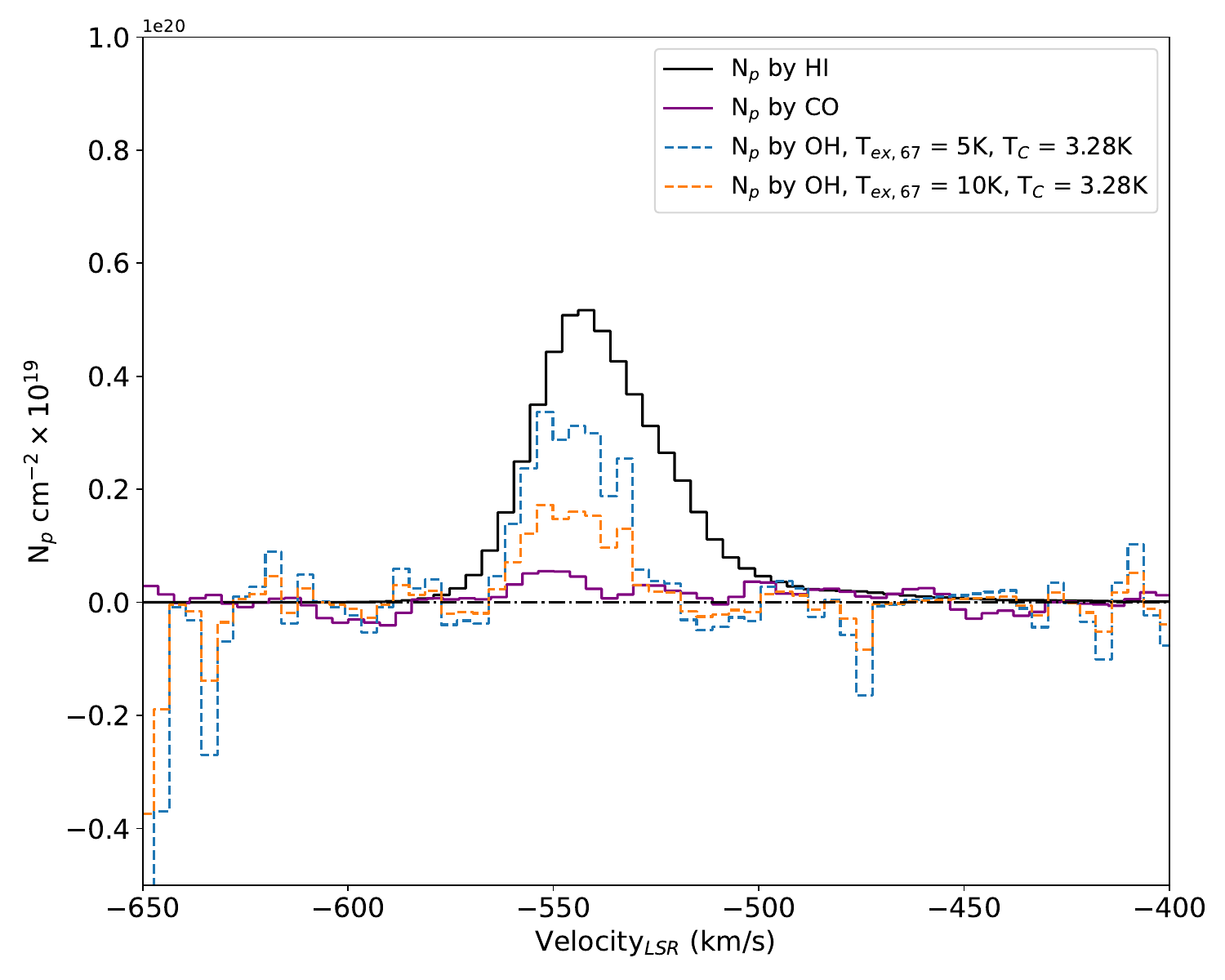}
    \caption{HI, OH and CO for the sightline discussed in this paper, transformed into proton (atomic hydrogen for HI, molecular hydrogen for OH, CO) column density per channel (N$_{p}$). We use the Galactic X(CO) value suggested from \citet{Bolatto2013TheFactor}. The abundance ratio of $N(\mathrm{OH})/N(\mathrm{{H_{2}}})$ = 1 $\times$ 10$^{-7}$ was used to convert the OH 1667 line to a proton column. This could also be labeled as hydrogen nuclei column density per channel. We present two choices of T$_{ex}$, where 5K is a somewhat standard choice, and 10K can be considered a lower limit to OH column densities, as most observed values of T$_{ex}$ for the 1667 MHz OH line fall between 4-10K with an average value around 4-5K.}
    \label{fig:proton}
\end{figure*}

\subsection{Comparing the Gas Mass Traced by HI, CO and OH}

Recent studies have suggested a correlation between faint optically thin HI and faint optically thin OH emission on large spatial scales \citep{Allen2012Faint5circ, Allen2013ERRATUM:97, Busch2021}. A corollary was discussed in \citet{Allen2012Faint5circ} from the data presented from that survey that related the main-beam temperature of HI and OH (1667 MHz) as T$_{mb}$(OH) = 1.5 $\times$ 10$^{-4}$ T$_{mb}$(HI), with a scatter of 3mK. This relation seems to hold for T$_{mb}$(HI) $<$ 60K, above which the HI saturates, but OH temperature continues to rise because it remains optically thin. As discussed in section \ref{sec:data}, this relation guided our exploratory search. The relationship predicts a peak OH signal of 2.6 $\pm$ 3 mK in the sightline without a detection, which is consistent with no signal, so perhaps this is not surprising. The sightline with the OH detection was chosen because the predicted peak signal here is 5.3 $\pm$ 3mK. In building further extragalactic OH searches, this relationship should be consulted and refined. It is especially unknown how this relationship differs with metallicity, which should be a topic of further research.

In order to compare the gas masses traced by HI, CO and OH, we must first adopt suitable conversion factors for each gas tracer. For OH, we need the abundance ratio of OH/H$_{2}$ to convert our $N$(OH) to $N$(H$_{2}$). In this diffuse molecular gas regime we can start by adopting a reasonable estimate for the value for the OH/H$_{2}$ abundance ratio. We adopt the median literature value of 1 $\times$ 10$^{-7}$ \citep{liszt1996, Wiesemeyer2016Far-infraredClouds, Rugel2018OHTHOR, Jacob2019FingerprintingDeconvolution}. While chemical model calculations tend to produce abundance ratios of varying values over a range of cloud parameter space (8$\times10^{-9}$ to 4$\times10^{-6}$), observations of the abundance ratio cluster fairly tightly around 1 $\times$ 10$^{-7}$ \citep{Nguyen2018Dust-GasISM}. Given the size of the resolution element, the median literature value for the ratio is likely a fair approximation because we are averaging over many different environments. This value for the abundance ratio appears to hold for $N(\mathrm{OH})$ values as low as $\sim$ a few $\times$ 10$^{12}$ cm$^{-2}$ \citep{Jacob2019FingerprintingDeconvolution}. Most of the literature values were derived for local, solar metallicity gas. The metallicity in M31 at the galactic radius discussed here ($\sim$ 13 kpc) is statistically consistent with solar metallicity \citep{Sanders2012THEPNe}. Note here also that we do not correct by the inclination of the galaxy, so values are beam-averaged along the line of sight.

As OH is almost always optically thin, we can directly calculate the column density and transform this into a column of H$_{2}$. Using the updated Cepheid distance to M31, $D = 761 \pm 11$ kpc \cite{Li2021ATelescope}. Then we can calculate the total mass in the enclosed beam via the equation:

\begin{equation}
    M_{\mathrm{H_{2}}} = N(\mathrm{H_{2}})\ A_{\mathrm{H_{2}}} \mu\ m_{\mathrm{H}}
\end{equation}

where $N(\mathrm{H_{2}})$ is the traced H$_{2}$ column. A$_{\mathrm{H_{2}}}$ is the physical area covered by the GBT beam (which is related to the GBT solid angle and distance to M31 by: A$_{H_{2}}$ = $\Omega\ d^{2}$), $\mu$ is the molecular weight which we set to 2 and for now ignore contribution to the mass by helium, and m$_{H}$ is the weight of the hydrogen atom.

The variation of T$_{ex}$ in diffuse clouds has been observationally constrained; the distribution of T$_{ex}$ as presented in \citet{Li2018WhereDMG} is a good starting assumption when we lack coincident absorption and emission profiles. In addition, there was the recent determination of T$_{ex}$ = 5.1 $\pm$ 0.2K in front of the W5 star forming region using the novel 'continuum background' method \citep{Engelke2018OHW5}. Since we detect emission, we know that T$_{ex}$ has to be higher than the continuum temperature in this region, 3.28K. We therefore take a range of T$_{ex}$ = 5 $\pm$ 0.5K as a sensible approximation. The approximation of T$_{ex}$ $\gg$ T$_{c}$ is also a suitable lower limit to the OH column density as the background correction factor goes to unity.

We calculate the inferred N(H$_{2}$) column density for this sightline from the CO data. The CO-derived N(H$_{2, CO}$) is 5.8 $\pm$ 0.4 $\times$ 10$^{19}$ using the typical value of X(CO) of $2\times10^{20}$ \citep{Bolatto2013TheFactor}.

The mass of $H_{2}$ traced by OH is then 1.1-1.6 $\times$ 10$^{7}$ M$_{\odot}$, with the range arising from the choice of T$_{ex}$ between 4.5-5.5 K. For comparison, the amount of mass traced by HI in this same sightline is 3.3 $\pm$ 0.01 $\times$ 10$^{7}$ M$_{\odot}$, the error in the HI calculation is statistical. These estimates suggest there is about 3 times more atomic mass than molecular mass in this sightline. If we were using CO data alone, the same estimate would suggest that there is 15 times more atomic mass than molecular. Transforming all gas tracers to physical units (proton column density per channel, or hydrogen nuclei per channel) is demonstrated in Fig.\ref{fig:proton}. This exercise gives a visualization on the range of ``CO-faint'' (we say ``faint'' here as opposed to ``dark'' because CO is detected) molecular gas that is traced by OH, which is dependent on the assumed T$_{ex}$.

\subsection{Calibration of OH/H$_{2}$ by Dust Emission on kpc Scales}

We can use dust data available on M31 to estimate the total neutral gas column density \citep{Nguyen2018Dust-GasISM} to calibrate the abundance ratio of OH/H$_{2}$. The procedure for calibrating the abundance ratio in this way is as follows: first we obtain an estimate for the dust mass in our observed region (M$_{d}$), apply a dust-to-gas mass ratio (M$_{d}$/M$_{H}$) to obtain an estimate of the total hydrogen gas mass, subtract the contribution of the gas mass by atomic hydrogen (as measured from the $N$(HI) data), and then assume the residual amount of gas mass is molecular. We use the dust models of dust emission from \citet{Draine2014AndromedasDust} to calculate this quantity. Usually, this calibration is done on much smaller spatial scales than we are discussing here \citep{Nguyen2018Dust-GasISM}, but since we are averaging over many different environments we may be averaging over dust systematics as well. The diameter of our resolution element is 1.67 kpc at the distance of M31 \citep{Li2021ATelescope}, resulting in an area of 2.2 kpc$^{2}$. We obtain the dust map shown in Fig.\ref{fig:coordinate} and measure the median dust surface mass density in an aperture the size of the GBT beam (7.6') at our OH detection coordinates. We correct the map to the updated Cepheid distance of $D = 761 \pm 11$ kpc \citep{Li2021ATelescope}, which introduces a small correction by a few percent.

The resulting median dust surface mass density in the aperture is 2.5 $\times$ 10$^{5}$ M$_{\odot}$/kpc$^{2}$. We use the M$_{d}$/M$_{H}$ prescription from Equation 8 in \citet{Draine2014AndromedasDust} for our distance from the the center of M31 of 13 kpc. We divide the dust surface mass density by this ratio (0.0079) to obtain the gas surface mass density and further divide by the proton mass to obtain N$_{H}$, the column density of hydrogen nuclei. This procedure results in a N$_{H}$ of 4 $\times$ 10$^{21}$ cm$^{-2}$. We then use the corresponding HI column density (uncorrected for optical depth effects, as the correction from \citet{Braun2009ADensity} seemed to overestimate HI as it was incompatible with the gas mass predicted by dust by an order of magnitude) from the \citet{Braun2009ADensity} HI map, as the dust map from \citet{Draine2014AndromedasDust} used this map to infer their dust models. The HI column density, N(HI), measured in the same aperture towards the OH detection coordinates is approximately 3 $\times$ 10$^{21}$ cm$^{-2}$. We then subtract this from the total hydrogen nuclei column density inferred from dust (N$_{H}$), and assume the rest of the hydrogen nuclei is molecular (H$_{2}$). This results in a N(H$_{2}$) of 5 $\times$ 10$^{20}$ cm$^{-2}$. Finally, we can calibrate \textbf{the} N(OH)/N(H$_{2}$) abundance ratio with this result. The corresponding values for T$_{ex}$ = 5, 10K are 1 $\times$ 10$^{-7}$ and 4 $\times$ 10$^{-8}$ respectively. These results are consistent with the range of literature values and seem to assure us our assumed value of 1 $\times$ 10$^{-7}$. We should also consider if there is a correction for opaque HI inside the beam, at large galactic radii the correction for optically thick HI is expected to be quite low\citep{Koch2021AComponent}, and optically thick HI has not been able to reconcile the observed mass of ``dark gas'' previously \citep{Murray2018OpticallyISM}. As a check, a generous 30$\%$ correction to the N(HI) used to subtract from the dust results in a N(H$_{2}$) of 7.6 $\times$ 10$^{19}$ cm$^{-2}$ and a abundance ratio of 7 $\times$ 10$^{-8}$, which is still consistent with literature values of the abundance ratio.

\section{Conclusion} \label{sec:conclusion}

We have detected statistically significant extragalactic emission from faint, thermal (non-maser, LTE-like) interstellar OH for the first time. The emission is detected in the southern disk of M31 using ultra sensitive GBT observations towards one sightline aligned with the major axis, $\sim$ 62' from the nucleus of the galaxy. A non-detection is also presented which suggests that the detection presented is not an otherwise unknown instrumental artifact. The velocity structure of the OH follows closely that of the narrow HI component and broad faint CO, implying the existence of molecular gas on the same spatial scale as the atomic gas traced by a portion of the HI. We derived Gaussian line profile parameters to the observed OH and HI data. We also calculated the profile integrals of the OH 1667 and 1665 line and found that they are in the LTE ratio. We compared the OH data with archival CO data from \citep{Dame1993}. The corresponding estimates for H$_{2}$ column density traced by CO is 5.8 $\pm$ 0.4$\times$ 10$^{19}$ using the standard X(CO) conversion with the estimated 30\% error \citep{Bolatto2013TheFactor}. OH appears to trace 140\% more than CO in the same sightoine. When we use dust as a total neutral gas column tracer we calibrate the OH/H$_{2}$ abundance ratio between 1 $\times$ 10$^{-7}$ and 4 $\times$ 10$^{-8}$ respectively, the range arising from the choice of excitation temperature.

Single dish radio telescopes with ample observing time at cm wavelengths (cm observations can normally be completed in any weather and any time of day) could systematically map the distribution of the 18cm OH emission in local group galaxies at large galactic radii, where the continuum temperature is only slightly above $T_{CMB}$. On the other hand, we may also expect to map out the absorption of 18cm OH towards the inner disks of these galaxies due to increased synchotron emission, which can be interpreted under slightly different assumptions. While CO will undoubtedly remain our brightest general tracer of H$_{2}$ in the Galaxy and beyond, mapping OH will drastically improve our understanding of the distribution and mass of the ``CO-dark'' gas phase. Single-beam L-band (cm) instruments would potentially require a prohibitive amount of time to survey local group galaxies at the 18 OH. New multi-beam phased array feeds \citep[e.g.][]{Roshi2018PerformanceTelescope, Pingel2021CommissioningFLAG} will drastically improve mapping speeds for large single-dish telescopes by orders of magnitude. In the near future, these instruments will be able to map the distribution of 18cm OH to trace the ``CO-dark'' gas, as long as sufficient sensitivities are achieved (T$_{rms}$ $<$ 1mK). The amount of molecular gas revealed by these new OH observations are significant, between 140\% more than the CO in this sightline. These estimates are similar in magnitude to the ``CO-dark'' gas estimates from $\gamma$-ray observations \citep{Grenier2005UnveilingNeighborhood} in the Milky Way's halo. It is still unclear if such a correction could be appropriately applied to the current CO maps of M31 and more observations are warranted. Mapping the distribution of OH would complement existing CO maps in the local group to help accurately understand the total accounting of the molecular phase of the ISM.

\begin{acknowledgments}
Michael P. Busch is supported by an NSF Astronomy and Postdoctoral Fellowship under award AST-2202373. We would like to thank Tom Dame for delivering the M31 CFA CO data in a digital format in private communication. We would also like to thank long-time collaborators Dave E. Hogg and Philip Engelke for sharing insightful scientific conversations and discourse. We want to thank and remember the late Ronald J. Allen for their mentorship and for the initial conception of this experiment. Thanks to the Allen family for their kindness and support. We would like to thank Karin Sandstrom and Josh Peek for comments, suggestions and guidance. We would also like to thank Joanne Dawson, Anita Petzler, David Neufeld, Michael Rugel, and Arshia Jacob for lively conversations about the OH molecule, its excitation mechanisms and how to observe it. The authors acknowledge Interstellar Institute's programs ``II4'', ``II5'', and ``II6'' and the Paris-Saclay University’s Institut Pascal for hosting discussions that nourished the development of the ideas behind this work.
\end{acknowledgments}

%

\vspace{5mm}
\facilities{Robert C. Byrd Green Bank Telescope}

\software{astropy \citep{TheAstropyCollaboration2013Astropy:Astronomy}; scipy \citep{Virtanen2020SciPyPython}, lmfit \citep{Newville2014LMFIT:Python}}

\bibliography{Mendeley}{}
\bibliographystyle{aasjournal}

\appendix

\section{Archival CO Data}

In this appendix we present the archival CO data, both the spectrum of the original and reprocessed data, and the approximate location of the extraction from the original CO data cube, which in the coordinates used of \cite{Dame1993} are approximately at $X$,$Y$ = [-62.4', 1.8'], where these are offsets from the center of M31. These offsets were calculated by calculating the offset from the Arp Outer Arm S5 to the OH detection source coordinates. The location of the Arp Outer Arm S5 in CO at $X$,$Y$ = [-63', 0'] (\citet{Dame1993}, Table 1), is calculated with a directional offset of 63' to the J2000 center of M31 at a position angle of $\theta$ = 37.7$^{\circ}$. There is an offset of approximately $\sim$ 2' to the NW as defined in \citet{Dame1993}. While offset from the CO peak at the Arp Outer Arm S5, it still lies within the GBT beam. The results from the Gaussian fit to the CO data are reported in Table \ref{table:fits}.

\begin{figure}[b]
    \centering
    \includegraphics[width=0.8\textwidth]{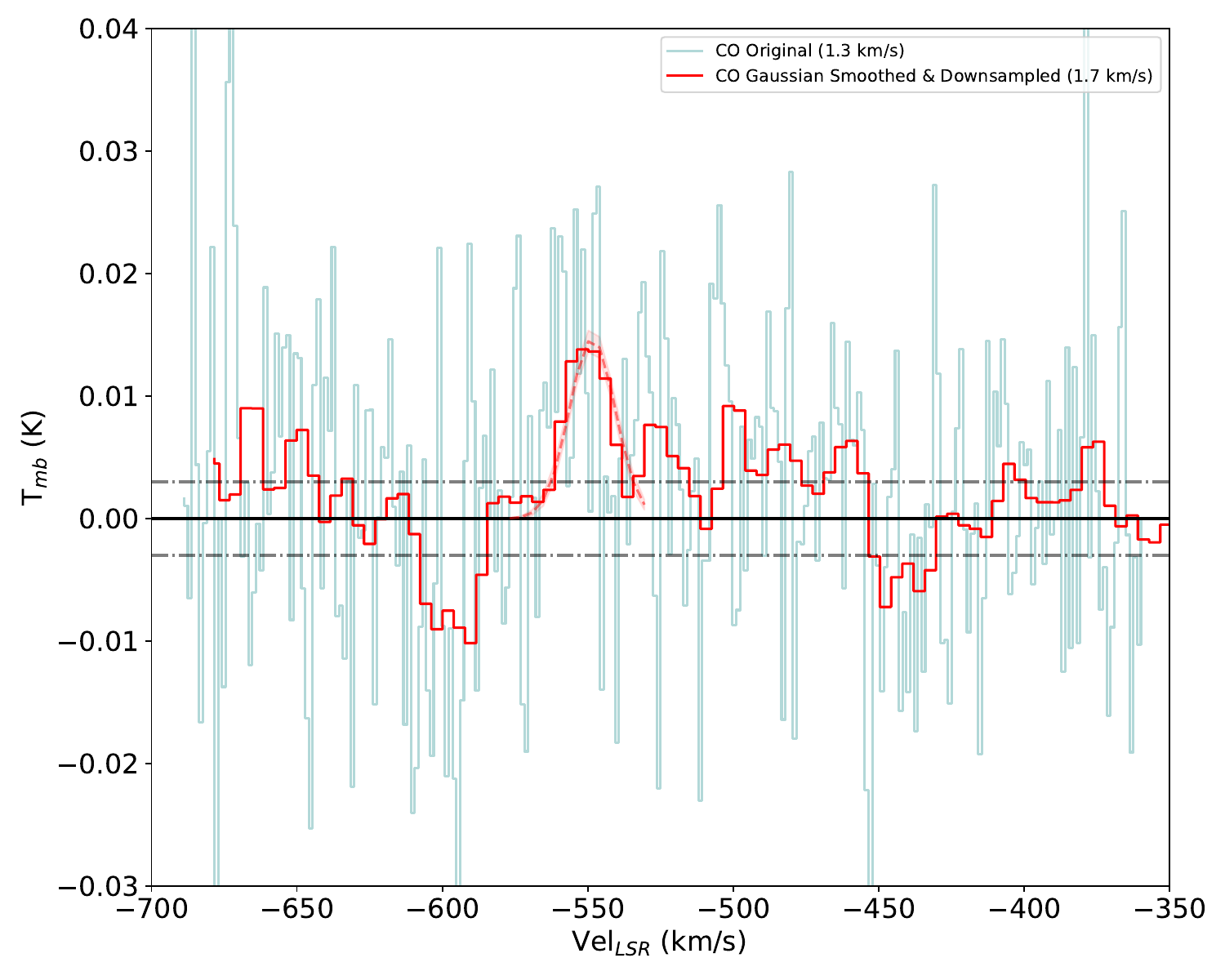}
    \caption{The original CO spectrum extracted from the M31 CFA CO data cube \citep{Dame1993} with the reprocessed spectrum overplotted. The dashed horizontal black lines represent the 1-$\sigma$ rms of noise of 0.003K. The dashed red line is a Gaussian fit to the reprocessed data, the Gaussian fit parameters are reported in Table. \ref{table:fits}.}
    \label{fig:CO}
\end{figure}

\begin{figure}
    \centering
    \includegraphics[width=\textwidth]{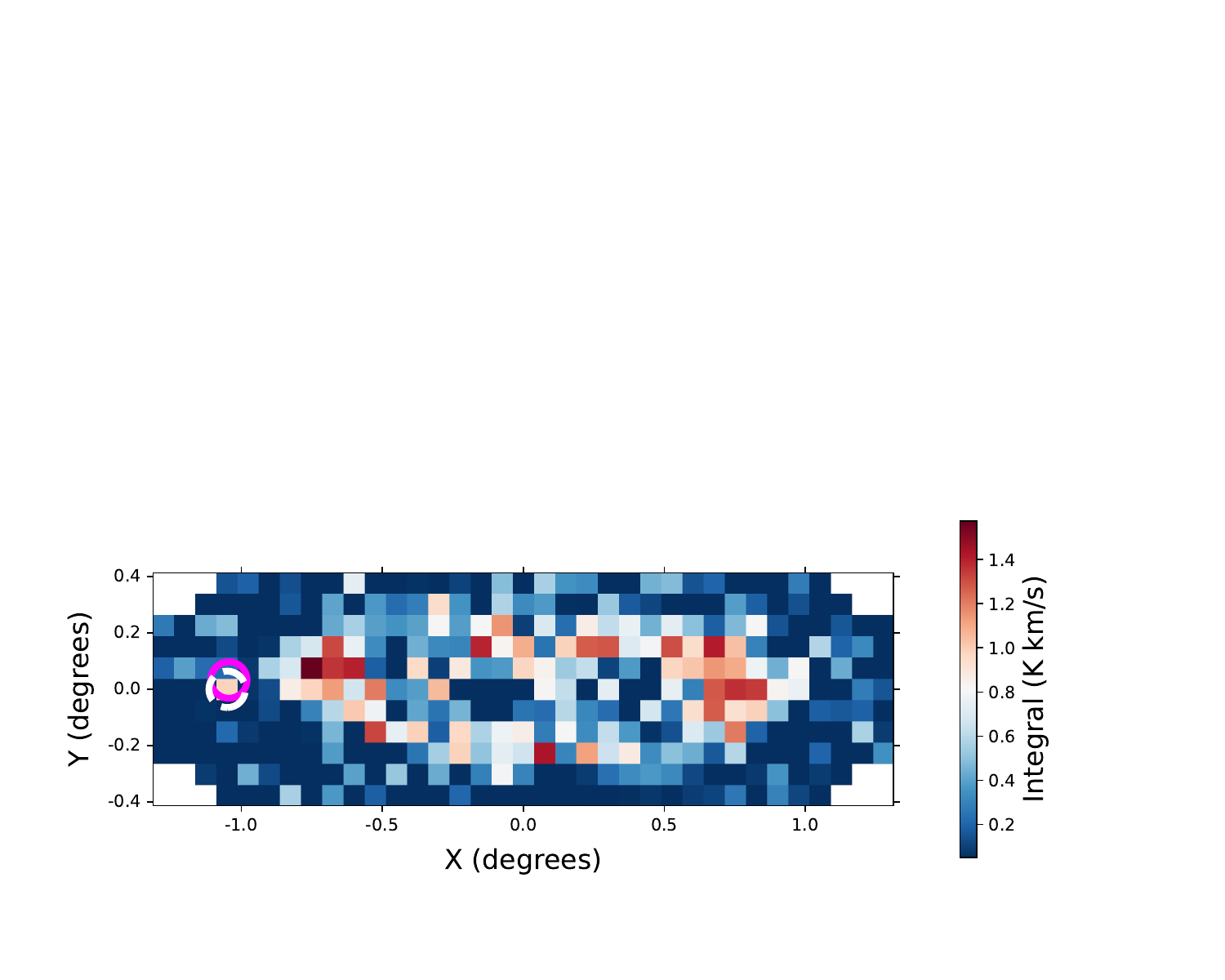}
    \caption{The integrated CO map from \citet{Dame1993} with the position of the OH detection sightline highlighted with the magenta aperture. The white aperture is the location of the Arp Outer Arm S5, previously discovered in CO, which coincides with our OH detection. The aperture is approximately the beamsize of the GBT (7.6'), where the CO spectrum from Fig. \ref{fig:CO} is extracted from. The $X$ and $Y$ are rectangular offsets from $\alpha$ = $0^{h}40^{m}$, $\delta$ = $+41^{\circ}$ (1950), with $X$ increasing toward the northeast along the major axis of M31 at a position angle of 37.7$^{\circ}$.}
    \label{fig:COmap}
\end{figure}

\end{document}